\documentclass[pra,twocolumn]{revtex4}
\usepackage{amsmath}
\usepackage{amsfonts}
\usepackage{amssymb}
\usepackage{graphicx}
\usepackage{sidecap}
\usepackage{caption,subcaption}
\usepackage{bm}

\usepackage{graphicx}

\newcommand{\Rmn}[1]{\uppercase\expandafter{\romannumeral #1}}

\allowdisplaybreaks
\makeindex

\begin{document}
\title{The Pauli principle in collective motion: Reimagining and
 reinterpreting Cooper pairs, the Fermi sea,
Pauli blocking and superfluidity.} 
\author{D. K. Watson \\
University of Oklahoma \\ Homer L. Dodge Department of Physics and Astronomy \\
Norman, OK 73019}
\date{\today}

\begin{abstract}
Typically visualized from an independent particle viewpoint,  the Pauli 
principle's role in collective motion is analyzed leading to 
a reimagination of the microscopic dynamics underlying 
superfluidity/superconductivity and a 
reinterpretation  of several interrelated phenomena: 
Cooper pairs, the Fermi sea, and Pauli blocking.  
The current approach, symmetry-invariant perturbation theory is a 
first principles method with no adjustable
parameters.  An adiabatic evolution is employed to transfer the well-known 
Pauli restrictions for identical,
independent particles with two spin values 
to restrictions on the collective modes of an ensemble 
of ``spin up'' ``spin down''
particles. The collective modes, 
analytic N-body normal modes, are 
obtained from a group theoretic exact solution of the first-order
equations.  
Cooper pairing is reinterpreted not as the pairing 
of two fermions with total zero momentum, but as
the convergence of the momentum of the entire ensemble
to two values,  $+k$ and $-k$, 
as the particles in the normal mode move back and forth with a single
 frequency and phase.
The Fermi sea and Pauli blocking, commonly described using 
independent fermions that occupy lower states to create a ``sea'' in 
energy space and block occupation is redescribed as a
collective energy phenomena of the entire ensemble.  
Superfluidity, which has always been viewed as a collective phenomena 
as Cooper pairs 
are assumed to condense into a macroscopic occupation of a single lowest
state, is now reimagined without two-body pairing in real space,
but as a macroscopic occupation 
of a low-energy phonon normal mode resulting in the convergence of 
the momentum to two equal and opposite values. 
The expected properties
of superfluidity including the rigidity of the wave function, interactions
between fermions in different pairs, convergence of the momentum
and the gap in the excitation spectrum are discussed.

\end{abstract}


\maketitle

\section{Introduction}

The Pauli principle plays a fundamental role in organizing
matter in our hadronic universe, providing stability 
as it intervenes in a wide range of
phenomena from the structure and dynamics of atoms to the physics of neutron 
stars\cite{dyson}. Restricting the 
permutation symmetry of indistinguishable particles and controlling
the occupation of identical fermions, it is responsible for 
the prevalence of degenerate Fermi systems
at all scales, forms the foundation of the periodic table and exists
at the core of quantum field theory.
Providing an effective repulsion that depends
 on particle statistics rather than 
interparticle interactions, the Pauli principle can dominate the
physical interactions particularly at ultralow temperatures 
and control the dynamics creating systems that
 exhibit universal behavior as found in
nuclear matter and in ultracold Fermi superfluids at
unitarity.

In a series of recent 
papers, the Pauli principle has been applied to ultracold superfluid Fermi 
gases\cite{inguscio,randeria,zwerger,strinati} without explicit 
antisymmetrization by using an adiabatic transition
between an independent particle regime and an interacting 
regime\cite{harmoniumpra,partition}.
This interacting regime is represented by normal modes which are the
first-order solutions of a perturbation series called symmetry-invariant
perturbation theory, SPT. 
The studies, initially at 
unitarity and more recently across the BCS to unitarity 
transition\cite{prl,emergence,jlow}, have taken advantage of the analytic 
form of the normal modes to study
the microscopic dynamics underlying these properties including superfluid
behavior. These dynamics are based on normal mode motions and thus 
differ from the conventional view that the particles in a 
superfluid form loosely bound bosonic pairs that condense into a macroscopic
occupation of the lowest 
state\cite{leggett1,leggett2,eagles,nozieres,jin1,jochim1,zwierlein1,zwierlein2,thomas1,salomon1,jin2,grimm1,hulet1}.  In addition to producing good agreement
with multiple experimental properties, normal mode dynamics
 offer a clear explanation 
for universal
behavior at unitarity that is lacking in two body approaches.

The success of this approach 
 using normal modes has demonstrated an interesting result: two-body pairing
is not necessary to describe superfluidity for ultracold Fermi gases.
Normal modes provide an alternative route to a macroscopic wave function
with phase coherence over the entire ensemble without the need for two-body pairing 
in real space, i.e. bosonic particles. This
refocuses attention on the importance of inter-pair correlations 
which are due to the Pauli principle and have always been recognized as 
crucial for an accurate
description of superconductivity/superfluidity.
Multiple studies in the field of condensed matter have questioned the
underlying microscopic basis of 
superconductivity/superfluidity\cite{combescot1,combescot2,combescot3,wietek,
dukelsky1,dukelsky2,
dukelsky3,dukelsky4,dukelsky5,solis}.  
In particular,
recent investigations\cite{combescot1,combescot2,combescot3}
 using an exact solution of the BCS Hamiltonian 
demonstrated that the composite bosons, i.e. Cooper pairs,
do not all condense into the same state as assumed in the original BCS
ansatz\cite{bcs}. This result suggests
that the Cooper pairs are not simple
bosons, and thus the microscopic basis of 
superconductivity/superfluidity may not be well understood, opening up the 
possibility that an alternative
microscopic description could underlie this important phenomenon. 

Based on these findings and the analysis of the
microscopic basis underlying superfluidity using normal modes, 
the goal of this paper is to reinterpret some of the seminal ideas
behind conventional approaches to superfluidity, to reimagine the
microscopic basis underlying superfluidity and to elucidate the role
of the Pauli principle in the emergence and stability of collective
behavior.

The Pauli principle
dominates the inter-pair interactions in the 
BCS ansatz\cite{combescot1,combescot2,combescot3}, 
and is critical to 
producing important properties of superfluidity/superconductivity
including an energy gap in the excitation spectrum, the rigidity of the
superfluid wave function that yields the Meissner effect, and the 
vanishing resistance to current flow.
It is interesting that early work did not assume simple 
two-body pairing in real space.
The highly successful BCS theory proposed in 1957\cite{bcs} 
assumes that the fermions are
 paired in momentum space with $+k$ and $-k$ values, i.e. zero-momentum
states.
As stated in the 1957 paper, 
the BCS wave function describes 
the ``coherence of large numbers of electrons,'' 
but does not propose that fermion pairs are
localized into
pseudomolecules that transition as in Bose-Einstein
condensation\cite{bcs}. As suggested by London in
1950, a superconductor is a ``quantum structure on a macroscopic scale...
a kind of solidification or condensation of the average momentum distribution'' of the electrons. ``It would not be due to distinct electrons at separate
places having the same momentum'', but ``it would arise from wave packets of
wide extension in space assigning the same local momentum to the entire
superconductor''\cite{london1}.  These early concepts of 
superfluidity/superconductivity as well as the seminal properties: 
pairing in momentum space, the long-range order over macroscopic distances,
 a ``rigidity'' of the wave function, and the gap in the excitation 
spectrum\cite{london2,schrieffer} are
naturally manifested in a normal mode picture of superfluidity.

Unlike many of the approaches to superconductivity 
which use the grand canonical ensemble to
avoid dealing directly with the Pauli exclusion principle, 
including the original variational procedure of the BCS ansatz, 
the Bogoliubov procedure\cite{bogoliubov1,bogoliubov2,tolmachev}
and Gorkov's Green's function method\cite{gorkov}, the current
SPT method applies the 
Pauli principle directly at first order using relationships between
independent particle quantum numbers and interacting normal mode quantum
numbers.
The SPT approach not
only avoids numerically intensive methods of enforcing antisymmetry, but
also reveals the role of the Pauli principle in a straightforward, transparent
way as it dominates the physics, determining the type, energy and degeneracy
i.e. the spectrum of normal modes  and thus the dynamics.

The microscopic basis underlying this method has been thoroughly analyzed in two
studies\cite{annphys,prafreq} taking advantage of the analytic forms
of these first-order N-body normal mode coordinates and their frequencies.
These studies track the emergence of collective behavior across the
BCS to unitarity transition as individual particles respond to 
increasing interparticle interactions and adopt collective
behavior that results in the particles moving in lockstep 
with a single frequency and phase as the average momentum 
converges to two equal and opposite values. 
The microscopic behavior that produces
universality, which is unexplained
 in two-body approaches, is clearly revealed by this 
normal mode picture\cite{prafreq}.

The present paper is an attempt to offer an alternative 
microscopic understanding of superfluidity accompanied
by a reinterpretation of the inter-related phenomena of Pauli blocking, the 
Fermi sea and Cooper pairing. The concepts of the Fermi sea and 
Pauli blocking, conventionally viewed from an independent particle
perspective, have had a large impact on our understanding of degenerate 
Fermi systems which 
exist at all scales throughout the universe. These phenomena, 
including atomic fermions
in cold Fermi gases; electrons in metals, white dwarfs, and heavy atoms; and
nucleons in atomic nuclei and neutron stars are now reimagined from a
collective viewpoint.

\section{Background}

Normal mode behavior pervades our universe at all energy and length 
scales manifesting the widespread existence of vibrational forces in nature
that occur in different media and across many 
orders of magnitude\cite{NM1,NM2,NM3,NM4,NM5,NM6,NM7,NM8,NM9,NM10,
NM11,NM12,NM13,NM14,NM15}. The
 many-body effects of large ensembles are encapsulated into simple dynamic 
motions coupling behavior into stable collective 
motion in which two-body pairing in real space is irrelevant. 
In the SPT approach, these collective motions have some measure of stability
since they are eigenfunctions of an  
approximate Hamiltonian\cite{annphys}.  
They offer
analytic solutions to many-body problems 
and physical insight into the microscopic dynamics underlying various
phenomena.

During the last eight years, the ability of normal modes to describe 
superfluidity  has been investigated for ultracold Fermi gases. Properties
at unitarity as well as across the BCS to unitarity transition were
obtained in 
close agreement with experiment  
including ground state energies\cite{prl}, critical temperatures\cite{jlow}, 
excitation frequencies\cite{prafreq,jlow}  
thermodynamic entropies and energies\cite{emergence,jlow}
as well as the lambda transition in the 
specific heat\cite{emergence}, a well known signature of the onset of
superfluidity.
At ultracold temperatures, only the lowest two types of normal
mode frequencies are relevant,  gapless phonon modes with 
extremely low frequencies and  particle-hole excitation modes. 
The particle-hole/single-particle excitation normal mode describes the excitation of a 
single particle out of the synced motion of the
many-body phonon mode, rather than the breakup of an individual pair of 
fermions.

The analytic frequencies of the normal modes 
were studied from the BCS regime to unitarity\cite{prafreq}
revealing the emergence of excitation gaps that increased from extremely
small gaps deep in the BCS regime to a maximum at unitarity as observed
in experiments. The
microscopic dynamics responsible for the emergence of these gaps were
investigated using the analytic forms of the normal mode functions.
In addition, a microscopic basis for the universal behavior at unitarity 
was proposed based on the synchronized normal mode motions 
which minimize 
interparticle interactions creating phenomena that are insensitive
to these interactions and thus have universal behavior.
 
The evolution in character of the $N$-body normal mode coordinates
was tracked by taking advantage of their rather simple analytic forms
(See Ref.~\cite{emergence} after Eq. (31).)
that take into account the complicated interplay of the particles as
they interact and cooperate to create collective
macroscopic motion\cite{annphys}. The symmetry in the Hamiltonian is the primary driver
of this behavior, thus these behaviors could be relevant for 
a wide range of
phenomena at different scales that are dominated by the same symmetry.

\section{Symmetry-Invariant Perturbation Theory: A Group Theoretic and Graphical Approach }\label{sec:SPT}

\subsection{Overview}

Symmetry-invariant perturbation theory is a first-principle method 
with no adjustable parameters that employs group theory
and graphical techniques to avoid the intensive numerical work 
typical in conventional many-body methods. If higher-order terms are small,
the first-order normal
mode solutions offer physical insight into the underlying 
dynamics\cite{FGpaper,energy,paperI,JMPpaper,laingdensity,test,toth}.
The perturbation parameter in this approach 
is the inverse dimensionality of space. 
Using $1/D$ or $1/N$ expansions to study physical systems was originally
developed by t'Hooft in quantum chromodynamics\cite{t'hooft}, 
 and subsequently used by Wilson\cite{kenwilson}
in condensed matter to calculate
 critical exponents for $D=3$ phase transitions
starting from exact values at $D=4$. These techniques have now been
used in many fields
of physics from atomic and molecular 
physics\cite{survey1,survey3,many1,many2,many3,many4,atomic1,
highfields2,HatomEM2,
quasi2,atomic2,atomic3,
atomic4,atomic5,atomic6,
atomic7,atomic8,atomic9,atomic10}, and 
condensed matter\cite{kenwilson,condensed1,condensed2,condensed3}, 
to quantum field 
theory\cite{whitten,quantumfield1,quantumfield2,quantumfield3,quantumfield4,
quantumfield5,survey6}, among others\cite{BEC1,BEC2,veillette,nishida1,
nishida2,nishida3,survey4,
relativistic1,relativistic2,relativistic3,relativistic4,nuclear1,nuclear2,statistical1,statistical2,whitten}.

The SPT formalism was developed to handle
the large ensembles at ultracold temperatures of interest in the atomic
physics/condensed matter communities  and was initially
applied to bosonic systems\cite{FGpaper,energy,paperI,JMPpaper,laingdensity}.
More recently, this formalism was extended 
to ultracold Fermi gases\cite{prl,emergence} which are subject to the constraints of the 
Pauli principle\cite{prl,emergence,harmoniumpra,partition}. 
Currently, this method 
has been 
formulated through first order for $L=0$ 
systems in three dimensions that are confined by spherically-symmetric potentials and 
have completely general interaction potentials. 
The SPT approach employs symmetry to attack the $N$-scaling
problem\cite{FGpaper,JMPpaper,paperI,liu,montina2008}, 
rearranging the work required for an
exact solution of the quantum $N$-body problem
 so the exponential scaling depends
on the order of the series, not the value of $N$ which becomes a simple parameter. 
Maximal symmetry is accessed by formulating a 
perturbation series about a large-dimension
configuration whose point group is isomorphic to the symmetric group $S_N$\,, 
and then evaluating the series for $D=3$.
The perturbation terms are evaluated for this
large dimension structure yielding terms that are invariant 
under the $N!$ operations of the $S_N$ point group.
This strategy allows the work at each order that scales exponentially to be
extracted as a pure mathematical problem\cite{rearrangeprl,complexity}.
In principle, this problem can be solved exactly using group theoretic 
methods, and 
then saved\cite{epaps}, with a significant reduction in the
 numerical cost of attacking a problem
 with a new interaction potential.

Even the lowest SPT perturbation order contains
beyond-mean-field effects that have produced excellent
results at first order\cite{energy,prl,emergence} as seen in 
earlier dimensional approaches\cite{herschbach1,herschbach2,loeser,kais1,kais2}.
This formalism has also been tested on an exactly solvable, fully-interacting
 model problem of harmonically-confined, harmonically-interacting 
particles\cite{test,toth,harmoniumpra,partition}. 
The agreement between the SPT wave function compared to the exact wave 
function is ten or more digits
verifying this general three-dimensional many-body formalism\cite{test} 
including the analytic formulas derived 
for the $N$-body normal mode coordinates and frequencies.

\subsection{The SPT formalism}

This section reviews the normal mode solutions, the application of the
Pauli principle in the selection of these modes and the determination
of their degeneracies. Detailed summaries of the SPT formalism can be found in 
Refs.~\cite{jlow,prafreq,emergence,annphys}

\subsubsection{Normal Modes}

The first-order perturbation equation is an harmonic equation which has 
been solved exactly, analytically for $N$ identical particles for both the 
normal mode frequencies\cite{FGpaper} and coordinates\cite{paperI} by
employing the FG method, a quantum chemistry method introduced by Wilson in 
1941\cite{wilson} to study the normal mode behavior of molecules\cite{dcw}. 
The solution involves five irreducible representations of
$S_N$\cite{hamermesh,WDC}, labelled   
${\bm 0^+, 0^-, 1^+, 1^-, 2}$\cite{FGpaper} where 
 the single normal mode of type ${\bm 0^+}$  
is a center of mass/symmetric bend motion;
the single normal mode of type ${\bm 0^-}$ is a 
breathing motion/symmetric stretch;
the $N-1$  type ${\bm 1^+}$ normal modes have particle-hole/single-particle
angular excitation behavior;  the $N-1$
   type ${\bm 1^-}$  modes exhibit particle-hole i.e.
single-particle radial excitation behavior;  and
the $N(N-3)/2$  
 type ${\bf 2}$ normal modes are phonon modes.  
These motions are analyzed in detail in
Ref.~\cite{annphys}. 
The energy through first order in $\delta = 1/D$ is:
\cite{FGpaper}
\begin{equation}
\overline{E} = \overline{E}_{\infty} + \delta \Biggl[
\sum_{\renewcommand{\arraystretch}{0}
\begin{array}[t]{r@{}l@{}c@{}l@{}l} \scriptstyle \mu = \{
  & \scriptstyle \bm{0}^\pm,\hspace{0.5ex}
  & \scriptstyle \bm{1}^\pm & , 
  &  \,\scriptstyle \bm{2}   \scriptstyle  \}
            \end{array}
            \renewcommand{\arraystretch}{1} }
(n_{\mu}+\frac{1}{2} d_{\mu})
\bar{\omega}_{\mu} \, + \, v_o \Biggr] \,, \label{eq:E1}
\end{equation}

\noindent where  $n_{\mu}$ is the total normal mode quanta
with frequency $\bar{\omega}_{\mu}$; 
 $\mu$ the normal mode label (${\bm 0^+, 0^-, 1^+, 1^-, 2}$),
$v_o$  a constant and multiplicities:
$d_{{\bm 0}^+} = 1, \hspace{1ex} d_{{\bm 0}^-} = 1,\;
d_{{\bm 1}^+} = N-1,\;  d_{{\bm 1}^-} = N-1,\;
d_{{\bm 2}} = N(N-3)/2$.

\subsubsection{Applying the Pauli Priniple} \label{subsec:pauli}

Eq.~(\ref{eq:E1}) gives the energy of the ground
state as well as the spectrum of excited states.
To determine the Pauli allowed states,  a correspondence is established between
the states of the non-interacting system, the three dimensional harmonic
oscillator $(V_{\mathtt{conf}}(r_i)=\frac{1}{2}m\omega_{ho}^2{r_i}^2)$ with 
radial and angular momentum quantum numbers $\nu_i$ and $l_i$, and the states 
of the interacting system identified by normal mode quantum numbers
$\vert n_{{\bm 0}^+},n_{{\bm 0}^-},n_{{\bm 1}^+},n_{{\bm 1}^-},n_{\bm 2}>$.  The
single-particle quantum numbers satisfy  $n_i = 2\nu_i + l_i$, 
where $n_i$ is the \textit{i}th particle energy level quanta defined by: 
$E=\sum_{i=1}^N\left[n_i  +\frac{3}{2}\right] \hbar\omega_{ho} =
\sum_{i=1}^N \left[(2\nu_i + l_i) +\frac{3}{2}\right] \hbar\omega_{ho}$. 
The constraints due to the Pauli principle are known for the harmonic oscillator 
states.  These constraints can be transferred to the normal mode
representation by taking a double limit
  $D\to\infty$, $\omega_{ho}\to\infty$ where both
representations are valid.  Since the radial and angular
quantum numbers separate at this double limit,
two conditions result\cite{prl,harmoniumpra}:
\begin{equation} 
\renewcommand{\arraystretch}{1} 
\label{eq:quanta}
2 n_{{\bm 0}^-} + 2 n_{{\bm 1}^-} =   \sum_{i=1}^N 2 \nu_i \, ,\,\,\,
2 n_{{\bm 0}^+} + 2 n_{{\bm 1}^+} + 2 n_{\bm 2} = \sum_{i=1}^N  l_i  \,
\renewcommand{\arraystretch}{1}
\end{equation}

\noindent Eqs.~(\ref{eq:quanta}) define a set of possible normal mode states
$\vert n_{{\bm 0}^+},n_{{\bm 0}^-},n_{{\bm 1}^+},n_{{\bm 1}^-},n_{\bm 2}>$
 consistent with the Pauli constraints
from the set of 
 harmonic oscillator configurations.
At the non-interacting
 $\omega_{ho}\rightarrow \infty$ limit, additional 
harmonic oscillator quanta, $\nu_i$ 
and $l_i$, 
are, of course, required by the Pauli principle as fermions fill the
harmonic oscillator levels. This corresponds to additional
 normal mode quanta required to obey the Pauli principle in the interacting regime.
The quanta are chosen to yield
 the lowest  energy for each excited state. For ultracold
temperatures, this results 
in occupation in $n_{\bf 2}$, the phonon mode, and in $n_{{\bf 1}^-}$, 
a single-particle radial mode, which have the lowest angular and radial  frequencies 
respectively. The conditions are:
\begin{equation} \renewcommand{\arraystretch}{1} 
\label{eq:quanta3}
2 n_{{\bf 1}^-}  =  \sum_{i=1}^N 2 \nu_i ,\,\,\,\,\,\,\,
2 n_{\bf 2} = \sum_{i=1}^N  l_i \,. 
\renewcommand{\arraystretch}{1}
\end{equation}
This strategy is analogous to use of the non-interacting system by Landau
 in Fermi liquid theory to set up the correct Fermi statistics as interactions evolve 
adiabatically\cite{landau}.

\medskip

A brief summary of the previous work using the SPT approach to calculate
superfluid properties of ultracold Fermi gases
using normal modes is given in Appendix~\ref{app:SPTcalculations}.
These calculations are modest in numerical costs since the
work to obtain the $N$-body normal mode coordinates and frequencies 
has been accomplished analytically.  
The first order results have no adjustable parameters
and are the exact results of this SPT
perturbation theory containing some correlation effects. The
thermodynamic results are more challenging to obtain requiring
a large number of energy levels to converge the partition function
as the number of fermions and/or the temperature increases.
Appendix B contains a brief discussion of the information encoded in the
two sets of quantum numbers.

\section{The Fermi sea and Pauli blocking: a collective viewpoint.}

From an independent particle view, degenerate Fermi systems have all the
lowest energy states filled, with a ``Fermi surface'' dividing the filled
from the unfilled levels. This ``sea'' of fermions exists in energy space,
with the scale of energies defined by the Fermi energy
which is the largest occupied energy in the system. The role of the Fermi sea
is to Pauli-block states below the Fermi energy, thus the behavior of such
systems is dominated by the Pauli principle which determines 
their general structure through the filling of the states. 
Degeneracy pressure, a striking feature of these systems,
is due to the Pauli principle requiring the occupation of high energy levels
and is responsible for the stability of
degenerate Fermi systems and thus their prevalence in our universe. 
Pauli blocking in ultracold, dense Fermi atoms has recently been verified
in the laboratory in the suppression of scattered 
light\cite{kjaergaard,ketterle,ye}.

The current approach now defines the concept of a Fermi sea from a
collective viewpoint, assuming that the particles 
are in a collective mode allowed by the Pauli
principle.  The Pauli restrictions originate in the independent particle
picture, but are transferred to the collective picture through an adiabatic
evolution of the system to the collective mode as interparticle interactions
turn on. The occupations of the lowest states in the independent particle
picture are responsible for the restrictions on the number of quanta 
permitted in the collective motion of the ensemble. For ultracold
systems at $T=0$, only phonon modes are occupied so the Fermi sea 
of occupied independent states becomes an energy minimum of the phonon 
collective mode,
with lower energy phonon modes unoccupied, i.e. blocked from occupation
by the Pauli principle.
The Fermi energy in the independent view is the energy of the highest
occupied independent state, while in the collective view, the Fermi energy
is the energy of the lowest allowed phonon mode.  For ultracold temperatures
greater than zero, multiple closely-spaced phonon modes may be occupied.

\section{Cooper pairs: a critical concept}
The concept of Cooper pairs has been called one of the pillars of the
microscopic theory of superconductivity, a concept that opened up the
route to a successful theory that could explain the physical effects
of zero resistivity, the existence of a gap in the excitation spectrum,
and the Meissner effect among others. 

Since the original BCS theory was introduced, the concept of Cooper pairs 
has evolved to include
a more nuanced understanding of the role played by this pairing. Multiple
studies have called into question the idea of a simple pairing of two
fermions which then condense as bosonic entities into the lowest 
state\cite{combescot1,combescot2,combescot3,wietek,dukelsky1,dukelsky2,
dukelsky3,dukelsky4,dukelsky5,solis}.
Interestingly, the original BCS paper clearly differentiated the pairing of 
momenta in BCS theory
from a simple bosonic pairing, stating:
 ``Our pairs are not localized in this sense and our
transition is not analogous to a Bose-Einstein condensation''\cite{bcs}. 
 By restricting configurations in their calculations to
 pairs of states with $+k$ and $-k$, i.e. zero momentum states, 
a coherent lowering of the energy was obtained.  This was ``consonant''
with ``London's concept of a condensation in momentum''\cite{schrieffer}. 
Theoretically, the pairs are
created by two fermion creation operators which
 do not satisfy Bose statistics. This is essential to the success of the
theory which must include many-body effects to yield an energy gap 
and long-range order over macroscopic
distances.  As reviewed by Bardeen in his Nobel address, ``A theory involving 
a true many-body interaction between the electrons seemed to be required to 
account for superconductivity''\cite{bardeennobel}.

These early concepts are consistent with a phonon normal mode picture, 
a rigid macroscopic wave function with long
range order extending
over the entire ensemble. The condensation of the frequency to a single
value as the particles
adopt the collective motion of a normal mode 
results in the expected convergence of the momentum
to two values, $+k$ and $-k$, as the particles slosh back and forth
in lockstep.
This normal mode picture retains Cooper pairing, a concept critical to the 
development of BCS theory, not as a two body phenomena in real space, 
but rather as
 a many-body phenomenon that
consolidates the momentum of an ensemble to two equal and opposite values.
This is consistent with the early recognized need for
a fully interacting many-body wave function.

\section{The seminal properties of superfluidity as supported by
the microscopic dynamics of the normal mode picture.} 

\subsection{ ``Rigidity'' of the wave function} This property of 
superconductivity has been called ``a striking manifestation of a subtle
form of quantum rigidity on the macroscopic scale''\cite{pines}. It prevents
a moderate external magnetic field from modifying the wave function
 and is also responsible for the gap in the excitation spectrum. 
Normal modes naturally provide rigid harmonic motion with the
 particles moving in lockstep with the same frequency and phase.  
In the SPT formalism these synchronized, collisionless
motions are eigenfunctions of an approximate Hamiltonian 
and thus have some degree of stability. They provide
 simple, quantum macroscopic wave
functions with phase coherence over the entire ensemble.
The microscopic behavior of the particles in a normal mode as they execute
rigid, harmonic motions is explored in detail in  Ref.~\cite{annphys}.

At present, explicit antisymmetrized SPT many-body wave functions have not been
 obtained. However, the specific normal modes that contribute to the many-body
wave function for a given energy level are known from the Pauli principle,
thus the general character of the wave function may be known particularly at 
ultracold temperatures.

\subsection{Interactions between the fermions in different 
pairs due to the Pauli principle. } The interactions between the fermion 
constituents of different ``bosonic'' Cooper pairs in BCS theory
are due to the Pauli 
principle\cite{combescot1,combescot2,combescot3}. Thus
the fermions in the BCS ansatz play a dual role: creating
 composite bosons that are assumed to condense to the lowest state as in Bose
Einstein condensation; and simultaneously 
retaining their fermionic nature giving
rise to inter-pair interactions from the Pauli principle. This dual
role calls into question the importance of two-body pairing as the 
underlying microscopic dynamic compared to the
 many-body correlations which are critical
to the emergence of superconductivity/superfluidity.

The normal mode picture reimagines this as simply the collective motion
of $N$ interacting fermions 
in a macroscopic normal mode wave function.  This normal mode
function at ultracold temperatures is a very low energy phonon mode.
Thus it is not necessary to produce bosonic entities that condense to the
lowest state to produce a macroscopic quantum wave function, 
as fermions can occupy one or more of the closely
spaced phonon modes to form a quantum, macroscopic function with collective
behavior. The many-body synchronous motion is subject to the Pauli principle 
which controls the structure of the interactions
at all strengths along the BCS to unitarity transition. 
  These inter-pair interactions in 
conventional BCS approaches are known to be crucial to producing
superfluidity/superconductivity.  The 
condensation to the lowest state by forming Cooper pairs is reimagined
as fermions occupying a very low energy phonon normal mode 
selected by the Pauli 
principle or possibly multiple closely-spaced modes each with
 a single frequency, rigid motion, and paired values
 of the momentum.

\subsection{``Solidification or condensation of the average momentum 
distribution'';  arising  ``from wave packets of
wide extension in space assigning the same local momentum to the entire
superconductor'': the uncertainty principle}\label{subsec:momentum}  
These quotes from London\cite{london1,london2} are
manifested in the BCS ansatz by assuming that the particles pair into
$+k$ and $-k$ pairs that condense into a single lowest state due to their
bosonic nature. The normal mode picture assumes many-body collective motion of
 the fermions in a phonon/compressional
normal mode.  As the particles begin
to move in sync with a single frequency and phase, the
 spatial extent of the normal mode expands, while
 the single frequency of motion means that the average momentum of the fermions
is converging toward a single absolute value as predicted by London
and others. This convergence in momentum space and the corresponding
expansion of the wave packet in position space is expected from the
uncertainty principle.
The following paragraphs analyze the SPT normal mode microscopic dynamics
underlying this condensation.

\subparagraph{The SPT microscopic dynamics underlying the condensation 
of the momenta for a phonon normal mode.}

The motions of the individual particles in a phonon normal mode have been
analyzed in Ref.~\cite{prafreq}.  The phonon normal mode is a purely
angular mode with three types of interparticle angular displacements:
a single, dominant, angular displacement;  
$2(N-2)$ nearest neighbor angular motions that move in response
to the dominant angle, opening if the dominant angle is compressing
and vice versa;
$N(N-1)/2-2(N-1)-1$ interparticle angles with a very small 
displacement for values of $N \geq 10$. (For very low values of $N$, 
$4 \leq N \leq 6$, the normal mode behavior is 
qualitatively analogous to behavior of small molecules such as
ammonia or methane.)

As $N$ increases, the proportions of the three groups of
interparticle angles  
evolve quickly with the third type, 
increasing as $N^2$, 
quickly becoming the overwhelmingly largest group in
the ensemble. 
For example, for $N=12$ there are a total of $N(N-1)/2=66$ 
interparticle angles, with one dominant angle, 20 nearest neighbor 
angles and 45 angles with the smallest response in the third group. 
When $N=100$ with a total of 4950 interparticle angles: there is a single
 dominant
angle, 196 nearest neighbor angles and  4753 angles with a very small response. 

Thus as $N$ increases,  ones sees the emergence of
compressional or phonon motion in this $[N-2,2]$ sector as
the angular oscillations push the atoms together 
and pull them apart without changing the radial positions. 
This is a compressional stationary wave
i.e. a phonon oscillation which is consistent with the 
very low frequency i.e long wave length of this mode\cite{prafreq}
and the large zero-point energy(See Eq.~(\ref{eq:E1}).)

\subparagraph{Pairing of the momentum in the three types of displacements.}
As shown above, the three types of interparticle angles result in
three different angular displacements. The momentum of a particle 
undergoing a particular displacement
is proportional to the frequency, and also proportional to the displacement.  
Thus, although every
particle in a normal mode moves with the same frequency, the different
displacements of the three groups mean that the momenta of the three
types of particles will be different.  The third group of displacements
which are the overwhelmingly largest group are
small, and decrease as $N$ increases.

\subsection{Gap in the excitation spectrum}  During the early 1950's
increasing evidence appeared for an energy gap at the Fermi surface.
This motivated the BCS ansatz of assuming that only zero momentum
pairs contributed leading to a lowering of the energy of the lowest 
state\cite{schrieffer}. In the normal mode picture, there is a natural 
gap between
the phonon mode and the next higher mode which is a single particle
excitation mode that increases from extremely small 
in the weakly interacting
BCS regime to a maximum in the unitary regime\cite{prafreq}. 
This gap provides stability for superfluid behavior particularly
as it widens as unitarity is approached. It also
leads to a value for the first excited state of the ensemble as
determined by the Pauli principle and an estimate of the critical
temperature that agrees with both the BCS estimate in weaker interaction
regimes as well as more intensive T matrix calculations of $T_C$
near the strongly interacting unitary regime\cite{jlow}. 
Gaps also exist between the
other types of normal modes (See Fig. 2 in Ref.~\cite{prafreq}) 
that could provide stability for collective
behavior if techniques to prevent the transfer to other modes exist
or could be engineered.

\subsection{The role of the Pauli principle}

The studies  by Combescot and coworkers\cite{combescot1,combescot2,combescot3}
revealed the seminal role of the Pauli principle in BCS theory
in producing superconductivity. Specifically, they
 looked at the $N$ dependence
of the pair energy using an exact solution of the BCS Hamiltonian.
The exact ground state is found to be formally different 
from the BCS ansatz precisely due to explicit terms that originate
from the Pauli exclusion principle between the inter-pair fermion
components. The $N$ composite bosons, i.e. the Cooper pairs,
do not all condense into the same state as elementary bosons do in 
Bose Einstein condensation, but actually multiple states are
involved.   This result suggests that the microscopic dynamics 
of superfluidity/superconductivity
are not well understood.

In the SPT approach, the Pauli principle controls the dynamics of the
collective behavior of the ensemble. The filling of energy levels in the
noninteracting limit dictates the selection of the occupied
normal modes, including the energy, the degeneracy and 
radial/angular type.  
At ultracold temperatures greater than zero, more
than one of the closely spaced phonon modes could be occupied which
is consistent with the results from the 
exact solution of the 
BCS Hamiltonian.

\section{Discussion}

The normal mode picture of superfluidity is attractive for multiple
reasons.  It adheres to the early concepts proposed by London and
by Bardeen, Cooper and Schrieffer in the original BCS paper.  
Specifically, it offers a rigid, macroscopic wave function with
momentum pairing of the motion of entire ensemble.
It sets up a natural gap between the low energy phonon
mode and a radial single particle excitation mode that is a maximum at 
unitarity.  

The motivation for proposing two body pairing of fermions into a bosonic
Cooper pair was to achieve a macroscopic wave function by 
condensation into the lowest state. 
Normal modes offer an alternate route to a
macroscopic wave function as fermions engage in collective behavior 
occupying a single normal mode 
energy level at $T=0$ and very closely
spaced phonon energy levels at ultracold temperatures greater than zero. 
In addition, a normal mode picture offers a microscopic
basis for universal behavior that is lacking in two-body approaches.

The agreement with experiment for ultracold Fermi gases is quite good
for multiple properties including thermodynamic properties. 
 The perturbation does not involve the strength of the interaction so it can 
be applied across the transition from
the BCS regime to the unitary regime.   Unfortunately,
these neutral atom ensembles do not permit a test of the Meissner effect.
The results
are obtained without adjustable parameters and are the exact first-order
results of the SPT formalism.
The good agreement suggests that
the beyond-mean-field effects included at first order are sufficient to
produce observable results and that the higher order SPT terms which 
have been formulated but not implemented are small for these low
temperature systems. In these regimes, the normal coordinates and frequencies
which depend on the interparticle interactions are, in fact, 
beyond-mean-field ${\it analytic}$ solutions to a many-body Hamiltonian
that have produced results in agreement with experiment, in some cases
based on a single frequency value\cite{jlow}.

At higher temperatures, or for larger $N$ values, as well as for the
convergence of the radial excitation frequencies near unitarity, evidence
exists that higher order terms are becoming important. 
The thermodynamic results also have some uncertainty
since they depend on the convergence of the partition function using the
infinite spectrum of equally-spaced excited energy levels.  This 
convergence becomes increasingly difficult as $N$ increases 
and/or the temperature
increases past the critical temperature. Fortuitously for these cases, 
small values
of $N$ yield quite good results as seen in multiple other few-body
studies\cite{adhikari,hu6,hu7,hu8,blume1,blume2,levinsen,grining}.

This picture does not offer a mechanism for the 
two-body pairing that occurs beyond unitarity
as the ensemble transitions to molecules in the BEC regime. 
Since the formalism is
a first-principles approach, higher order terms will eventually break down
the normal mode picture and could yield results for the BEC regime that
agree with experiment, but it is not likely that they would 
offer a clear picture of the transition to this regime.


The group theoretic work underlying this approach is robust, taking
advantage of the symmetry at high dimensions to do the
``heavy lifting'' in obtaining exact solutions at first order that
contain beyond-mean-field effects.  
The subsequent implementation of the group theoretic results in a fully 
interacting
approach through first order retains this robustness, being heavily based on 
the manipulation of integers.

The analytic nature of these
first-order solutions has allowed the microscopic dynamics to be
studied in detail leading to an explanation for the universal behavior
at unitarity. As the interparticle interactions increase toward unitarity,
the synchronized motions of a normal mode become increasingly
correlated which minimizes the dependence on the details of these
interparticle interactions since the interparticle distances are becoming
fixed. Thus, the strong interactions paradoxically result
in minimal interparticle interactions due to the lockstep motion. 
This is particularly true for the angular
normal modes whose frequencies converge to integer multiples of the
trap frequency setting up a spectrum analogous to the independent
particle limit supporting universal dynamics.

The normal mode solutions set up natural gaps between all the modes
that could support different phases of collective behavior
if mechanisms to prevent the transfer of energy to
other modes exist or could be constructed allowing one of the modes
to dominate.
Temperature is likely to
play a role in controlling the dynamics of such structures, but other mechanisms
may supercede the effect of temperature to perhaps allow
a high temperature system to sustain stable collective behavior which could be
harnessed to bring the desired benefits of quantum engineering
 on a macroscopic scale.







\section{Summary.} \label{sec:summary}

In this paper, the role of the Pauli principle in the collective motion
of ultracold identical fermions has been analyzed leading to alternative
understandings of four interrelated phenonmena: Cooper pairs, the Fermi sea, 
Pauli blocking and superfluidity.  This analysis is based on a series
of papers that determined properties of ultracold Fermi systems
first at unitarity and then across the BCS to unitarity transtion. These
calculations used the Pauli principle at the non-interacting limit
to determine degeneracies for the interacting many-body levels and
applied the Pauli principle to select the allowed normal
modes.  
The determination of properties along the BCS to unitarity transition 
and the investigation of the underlying dynamics has 
yielded physically intuitive behavior that has been observed
in the laboratory. 
The microscopic dynamics are based on normal mode motions 
and thus differ from the  
accepted view that loosely bound pairs form
that decrease in size as a Feshbach resonance is tuned to strong 
interactions. 

Superfluidity, perhaps the most well-known example of collective
behavior, has been attributed to two-body mechanisms that lead to a
Bose-Einstein-like condensation to the lowest state to achieve a macroscopic
quantum wave function.  The current SPT approach 
has suggested that there is an alternative mechanism that results in
a macroscopic wave function, namely the ability of emerging normal
mode behavior to organize a 
many-body system
into a rigid, collective macroscopic wave function with a single frequency,
converged momentum,
and phase coherence over the entire ensemble.

\section{Conclusions.} \label{sec:conclusions}

The successful use of normal modes to determine multiple 
properties of ultracold Fermi gases in
 excellent agreement with both experiment and theory has demonstrated
that two-body pairing assumptions in real space are not necessary to describe
superfluidity for the ultracold Fermi gases. This suggests that reimagining
the underlying microscopic basis for superfluid behavior in multiple
regimes could be worthwhile. Correlations between the
pairs have
always been known to be critical to describing superfluid behavior.  These
inter-pair correlations refocus the physics away from the importance of
two-body pairing to a many-body picture. A macroscopic
wave function with phase coherence and
rigid motion is manifested in the current approach
by normal modes  making two-body pairing irrelevant.

Particle statistics are known to be 
powerful organizational, driving
forces in the emergence of collective states of matter
for both bosons and fermions 
with simple behavior emerging from the complexity of the microscopic
world.  
Despite the importance of the Pauli principle in a wide range of phenomena, 
 this fundamental tenet still lacks a simple explanation, 
as pointed out by many physicists including Pauli himself in his Nobel 
address, prompting
both experimental tests\cite{ramberg,VIP,dolgov,english}  
and theoretical attempts
to yield deeper insight\cite{kaplan,schilling}. 
In this paper, the Pauli principle's role in collective motion 
as documented in previous successful investigations of ultracold Fermi
gases has led to a reimagination and
reinterpretation of the seminal concepts of superfluidity. 

The Pauli principle has long been recognized as being critical to 
obtaining an accurate description of superfluidity. The current work 
is an attempt to define and perhaps enlarge that role.
Understanding the role of the Pauli principle in collective motion 
could have profound consequences for our understanding of the dynamics
that support the emergence and stability of superfluidity and other
organized behavior in our universe.


\section{Acknowledgments}
I am grateful to the National Science Foundation for financial support
under Grant No. PHY-2011384.

\bigskip

\appendix
\renewcommand{\theequation}{A\arabic{equation}}
\setcounter{equation}{0}

\section{Discussion of SPT calculations of superfluid properties
compared to experiment.}
\label{app:SPTcalculations}

\subsection{Overview}

During the last eight years, the normal mode picture of 
superfluidity has been
tested by determining various properties of ultracold Fermi gases
both at unitarity and across the BCS to
unitarity transition comparing to experimental data. Due to the ultracold
temperatures, these calculations use only the lowest angular 
frequency of the phonon normal mode and the lowest radial frequency of
a single particle radial excitation frequency.  These frequencies have yielded
properties for the ground state energies comparable to benchmark
Monte Carlo calculations for $N < 30$\cite{prl} and critical temperatures
from BCS to unitarity in good agreement with theoretical results\cite{jlow}. 
Thermodynamic quantities including entropies, 
energies and the lambda transition in the 
specific heat\cite{emergence} used the spectrum of excited states
for these two lowest normal modes obtaining close agreement with
experiment although the cusp for the specific heat was lower
than most results in the literature. The thermodynamic calculations
relied on the enforcement of the Pauli principle to select the
energy levels included in the partition function 
sum and to determine the degeneracies of these levels.

\subsection{Application: ultracold Fermi gases}

In the previous studies of ultracold Fermi gases,
an $N$-body ensemble of fermions is assumed with $N=N_1+N_2$ where  $N_1$ 
and $N_2$ are the number of ``spin up'' and
``spin down''  fermions respectively with $N_1 =N_2$ and $L=0$
symmetry. The fermions are confined by a spherically symmetric
harmonic potential with frequency $\omega_{ho}$.
An attractive square 
well potential is defined with a radius $R$ and potential depth parameter 
$\bar{V_0}$ in scaled units. $\bar{V_0}$ is varied from a value of 1.0 where the 
s-wave scattering length, $a_s$, is infinity to a value of zero 
 as the gas becomes weakly interacting in the BCS regime. 
The range of the potential is chosen so $R \ll a_{ho}$.
(See Ref.~\cite{prl} for more details.)

The full SPT many-body formalism is applied to a Fermi gas along the BCS to 
unitarity transition, defining internal displacement coordinates
and obtaining symmetry coordinates \cite{FGpaper,energy}.
The FG method\cite{dcw} is employed 
to obtain the $N$-body normal mode coordinates and their frequencies 
analytically. The Pauli principle is used with
 Eq.~(\ref{eq:E1}) to determine the ground state energy
as well as the  spectrum of excited states used in constructing the partition
function.

The canonical partition function is defined as: 
$Z = \sum_{j=0}^\infty g_j \exp(-E_j/T)$, where $E_j$ is a many-body energy,
$T$ is the temperature ($k_B=1$), and  $g_j$ 
is the degeneracy of $E_j$. (See Section~\ref{subsec:degen}.)
Quanta for the lowest values of the normal mode
frequencies are chosen 
to yield the lowest excited energy level. 
(See Eq.~\ref{eq:quanta3}.)

\subsection{Partitioning the energy and collecting statistics: Degeneracies}\label{subsec:degen}

Determining the degeneracies of a many-body energy level is a nontrivial task. 
Perhaps the
most straightforward way is to diagonalize the many-body Hamiltonian which 
is limited for exact results to
systems with less than ten particles\cite{booth}.  The Fermi-Dirac distribution 
represents a high-$N$
approximation and various other methods have been tried including 
recursive methods which are generally restricted to lower values of 
$N$\cite{borrmann,brosens1,brosens2}.

In the present method, the degeneracies of the interacting system 
are obtained from the corresponding noninteracting system since 
the degeneracies are independent of  the two-body interaction strength; 
i.e. the number of 
states of a degenerate level remains fixed as the interaction varies 
continuously. 
The many-body degeneracy for {\it each} partition of the energy into
specific $n_i$, $i=1,,,,,N$ so that 
$E=\sum_{i=1}^{N}\left[n_i +\frac{3}{2}\right] 
\hbar\omega_{ho}$ can be
calculated by determining all possible arrangements of the particles 
filling 
in a sublevel,
$\nu_i, l_i$.  
These sublevel degeneracies for this partition 
are collected and multiplied together
 yielding the degeneracy for
this partition.  For a given total many-body energy, $E_j$, 
there can be many different
partitions whose degeneracies must be calculated and then added together 
to determine the
total many-body degeneracy for this energy level.  
Each of the many-body energy levels in the partition function sum
must be partitioned in this way
among all $N$ particles and the correct many-body degeneracy, $g_j$
determined.

The degeneracy obtained in the
noninteracting limit can then be used for the interacting system with
the collective quantum numbers obtained using  Eqs.~(\ref{eq:quanta3}).
These collective quantum numbers, $ n_{\bf 2}$ and  $n_{{\bf 1}^-}$,
which contain information from enforcing the Pauli principle,
 are used in Eq.~(\ref{eq:E1}) to obtain the many-body energy of 
the {\it interacting} system.

\subsection{Behavior of the frequencies from BCS to unitarity}

The behavior of all five frequencies was studied in 
Ref.~\cite{prafreq} across the BCS to unitarity transition. The angular
frequencies limit to integer multiples of the trap frequency as
seen experimentally as the angular motion becomes rigid and not sensitive 
to the interparticle interactions. The radial frequencies both go through
a dip as unitarity is approached and then continue to increase.  Unfortunately,
these radial frequencies do not appear to be converged at unitarity possibly 
requiring higher order terms.
The dip seen in the SPT breathing radial frequency, $\omega_0^+$, has been
 observed in the lab\cite{thomas1,bartenstein,grimm1,grimm2} and verified in previous theoretical work\cite{tosi,zubarev,salasnich}.  
This feature has been attributed by some studies to the formation of 
Cooper pairs. The SPT analysis using the analytic forms for this frequency
show that the dip is due to
the competition between the centrifugal contributions of the Hamiltonian
and the increasing interparticle interactions which combine to
create a dip right before strong interactions dominate.
Microscopically, the particles are undergoing an increase in their correlated motion  
which minimizes the interparticle interactions and results
 in slower oscillations of the breathing mode, i.e. a downturn. Eventually the
increase in $\bar{V}_0$, i.e. the increased strength of the
 interparticle interactions results in more rapid oscillations i.e. 
an increase in the frequency as unitarity is 
approached. 
Although this radial frequency is too high to be relevant to creating
a superfluid, the agreement of this feature of a single, analytic SPT
frequency with experiment suggests 
that the beyond-mean-field effects 
incorporated at first order are sufficient to produce
effects observable in the laboratory.

\section{Quantum numbers and information}\label{app:quantumnumbers}

It is interesting to consider the information encoded in the single particle
quantum numbers of the three dimensional harmonic oscillator 
compared to the information held in the normal mode
quantum numbers in the interacting system.  For the non-interacting
system, the single particle quantum numbers, $n_i$, $\nu_i$ and $l_i$,
which are related by:  $n_i = 2\nu_i + l_i$,  hold information about the
individual energies of each of the $N$ particles as well as their
radial and angular behavior. As identical fermions fill the states of
the oscillator according to the Pauli principle,
these quantum numbers acquire information from the Pauli principle
due to these constraints
which are well known in the 
noninteracting limit. For ultralow temperatures, Eq.~\ref{eq:quanta3}
is used to determine the values of the angular 
phonon normal mode quantum number,
$ n_{\bf 2}$, and the single particle radial excitation normal mode quantum 
number, $n_{{\bf 1}^-}$. These new collective quantum numbers are the sum of $N$
independent particle quantum numbers and of course no longer contain information about  each
individual particle since $n_i$, $\nu_i$ and $l_i$ are no longer good quantum numbers.
Instead, the evolving Hamiltonian which is coupling the particles
into normal mode motions, now has new collective quantum numbers encoding relevant
information relating to operators that commute with the interacting
Hamiltonian.  These operators are the number operators, one for each
irreducible representation, i.e. for each of the five types of normal
modes. The normal mode quantum numbers, $ n_{\bf 2}$ and  $n_{{\bf 1}^-}$,
relevant for temperatures near $T=0$ hold information about the energy of the
entire ensemble as well as allowed radial and angular behavior.  Their 
values also contain information from the enforcement of the Pauli principle 
which has controlled the filling of the noninteracting energy levels
and thus the resulting values of these collective quantum numbers.

\end{document}